\documentclass[aps,pre,twocolumn]{revtex4}
\usepackage{epsfig}

\begin{document}

\title{Subcritical mirror structures in an anisotropic plasma}
\author{E.A. Kuznetsov$^{(a,b,c)}$\/\thanks{%
e-mail:kuznetso@itp.ac.ru}, T. Passot $^{(d)}$, V.P. Ruban $^{(c)}$ and P.L.
Sulem $^{(d)}$}
\affiliation{{\small \textit{$^{(a)}$P.N. Lebedev Physical Institute RAS, 53 Leninsky
Ave., 119991 Moscow, Russia\\
$^{(b)}$Space Research Institute RAS, 84/32 Profsoyuznaya str., 117997,
Moscow, Russia\\
$^{(c)}$ L.D. Landau Institute for Theoretical Physics RAS, 2 Kosygin str.,
119334 Moscow, Russia\\
$^{(d)}$Universit\'e de Nice-Sophia Antipolis, CNRS, Observatoire de la
C\^ote d'Azur, PB 4229, 06304 Nice Cedex 4, France}}}

\date{\today}

\begin{abstract}

Based on Grad-Shafranov-like equations, a gyrotropic plasma where the 
pressures in the static regime are
only functions of the amplitude of the local magnetic field is 
shown to be amenable to a variational principle
with a free energy density given by the parallel tension. 
This approach is used to demonstrate that small-
amplitude static holes constructed slightly below the mirror 
instability threshold identify with lump solitons
of KPII equation and turn out to be unstable. 
It is also shown that regularizing effects such as finite Larmor
radius corrections cannot be ignored in the description of 
large-amplitude mirror structures.

\end{abstract}


\maketitle

\section{Introduction}

Pressure-balanced structures are commonly 
observed in space plasmas. They are often associated
with the nonlinear saturation of the mirror instability
(MI) \cite{VedenovSagdeev, Gary} which, being of subcritical type 
\cite{KPS07a,KPS07b},
permits the persistence of non-zero solutions below
threshold \cite{Soucek, Genot}. Furthermore, near the MI threshold,
the dynamics of weakly nonlinear mirror modes are
governed by an asymptotic equation of gradient type
[5, 6]. This property implies an irreversible character of
the mirror modes behavior, associated with ion Landau
damping, where the free energy can only decrease in
time. In this framework, above the threshold, the
mirror modes have a blow-up behavior with a possible
saturation at an amplitude level comparable to that
of the ambient field. Below threshold, all stationary
(localized) structures were predicted to be unstable.
The main goal of this paper is to study 
stationary localized structures resulting from the balance of
magnetic and (both parallel and perpendicular) 
thermal pressures, whose simplest description is provided
by anisotropic MHD. Isotropic MHD equilibria are 
classically governed by the Grad-Shafranov (GS) equation
[7, 8, 9]. We here revisit this approach in the case of
anisotropic electron and ion fluids where the 
perpendicular and parallel pressures are given by equations
of state appropriate for the static character of the 
solutions. However, the MHD stationary equations, at
least in the two-dimensional geometry, turn out to be
ill-posed. As a consequence, these equations require
some regularization. As done in a similar context of 
pattern formation [10], an additional linear term involving a
square Laplacian is added. For nonlinear mirror modes,
regularization can originate from finite Larmor radius
(FLR) corrections, which are not retained in the present
analysis based on the drift kinetic equation (see, e.g.
[5, 6]). The paper is organized as follows. In Section 2,
the anisotropic Grad-Shafranov equations are revisited
when the gyrotropic pressures depend only on the local
magnetic field amplitude that, as shown in the 
forthcoming sections, is specific for nonlinear mirror modes.
In this case, as well known [11, 12, 13, 14], the 
parallel component of the equation is satisfied identically.
In Section 3, we show that in the two-dimensional 
geometry, the problem is expressed in a variational form
with a free energy given by the space integral of the
parallel tension. In Section 4, it is shown that the 
equations of state resulting from an adiabatic approximation
of the drift kinetic description, require a regularization
due to an overestimate of the contributions from the
particles with a large magnetic moment. We discuss
in particular the small-amplitude regime and show that
the pressure-balanced structures are then governed by
the KPII equation which possesses lump solutions. 
Numerical simulations reproduce these special structures,
that turn out to be unstable. Computation of stable so-
lutions lead to large-amplitude purely one-dimensional
solutions that appear to be sensitive to the regulariza-
tion process, an indication that the regime cannot be
captured by the drift kinetic approximation and that fi-
nite Larmor corrections and trapped particles are to be
retained. Section 5 is the conclusion.

\section{Anisotropic Grad-Shafranov equations}

{\bf Gyrotropic pressure balance}.
We start from the pressure balance equation for a static gyrotropic MHD
equilibrium%
\begin{equation}
0=-\nabla \cdot \mathbf{P}+\frac{1}{c}\left[ \mathbf{j}\times \mathbf{B}%
\right] ,  \label{MHD}
\end{equation}%
where the current $\mathbf{j}$ is defined from the Maxwell equation as%
\textbf{\ }$\mathbf{j}{=\frac{c}{4\pi }\nabla \times }\mathbf{B}$, and the
pressure tensor $\mathbf{P}$ is assumed to be gyrotropic. The solvability
conditions read $\mathbf{B}\cdot (\nabla \cdot \mathbf{P})\mathbf{=}0\mathbf{%
,}$ and $\mathbf{j}\cdot (\nabla \cdot \mathbf{P})\mathbf{=}0$.

In terms of the tension tensor $\Pi _{ij}=\Pi_{\perp }\left( \delta
_{ij}-b_{i}b_{j}\right) +\Pi_{\parallel}b_{i}b_{j}$, Eq. (\ref{MHD}) takes
the divergence form $\frac{\partial }{\partial x_{j}}\Pi _{ij}=0$. Here $%
\mathbf{b}=\mathbf{B}/B$ is the unit vector along magnetic field and $%
\Pi_{\perp } =p_{\perp }+B^{2}/(8\pi)$ and $\Pi_{\parallel } =p_{\parallel
}- B^{2}/(8\pi)$, 
where the perpendicular and parallel pressures $p_{\perp }=\Sigma _{\alpha
}p_{\perp \alpha }$ and $p_{\perp }=\Sigma _{\alpha }p_{\perp \alpha }$ are
the sum of the contributions of the various particle species $\alpha$. They
are expressed as $p_{\perp \alpha } = m_{\alpha }B^{2}\int \mu f_{\alpha
}dv_{\parallel }d\mu$ and $p_{\parallel \alpha } = m_{\alpha }B\int
v_{\parallel }^{2}f_{\alpha}dv_{\parallel }d\mu$, in terms of the
distribution functions $f_{\alpha }$, which satisfy the stationary drift
kinetic equations 
\begin{equation}
v_{\parallel }\nabla _{\parallel }f_{\alpha }-\Big( \mu \nabla _{\parallel
}B+\frac{e_{\alpha }}{m_{\alpha }}\nabla _{\parallel }\phi \Big) \frac{%
\partial f_{\alpha }}{\partial v_{\parallel }}=0,  \label{kineticEqs}
\end{equation}%
where $\nabla _{\parallel }=\mathbf{b}\cdot \nabla $ denotes the gradient
along magnetic field, $v_{\parallel }$ the parallel component of the
particle velocity, $\phi $ the electric potential, and $\mu =v_{\perp
}^{2}/\left( 2B\right) $ the adiabatic invariant (magnetic moment) which
plays the role of a parameter. These equations are supplemented by the
quasi-neutrality condition $\sum_{\alpha }e_{\alpha }B\int f_{\alpha
}dv_{\parallel }d\mu =0$, that allows one to eliminate the electric
potential.

We consider partial solutions of the stationary kinetic equations (\ref%
{kineticEqs}) which  are expressed in terms of two integrals of motion: the
energy of the particles $W_{\alpha }={v_{\parallel }^{2}}/{2}+\mu
B+(e_{\alpha }/m_{\alpha })\phi $ and their magnetic moment $\mu $. 
Besides these integrals, the solution can depend on the integral  which is
a  label to  each magnetic field line \cite{Grad1967}.  The choice  $%
f_{\alpha }=f_{\alpha }(W_{\alpha },\mu )$, as it will be shown in Section
3, can be matched with the solution found perturbatively for weakly
nonlinear mirror modes  \cite{KPS07a, KPS07b}.  In this case  the parallel
and perpendicular pressures for the individual species and also the total
pressures are functions of $B$ only. We write $p_{\perp \alpha }=p_{\perp
\alpha }(B)$ and $p_{\parallel \alpha }=p_{\parallel \alpha }(B)$. As seen
in the next subsection, this property plays a very central role in the
forthcoming analysis.

{\bf \protect\bigskip Identity along $\mathbf{B}$}.
The anisotropic pressure balance equation reads \cite{KPS12}%
\begin{eqnarray}
&&-\nabla \left( p_{\perp }+\frac{B^{2}}{8\pi }\right) +\left[ 1+\frac{4\pi 
}{B^{2}}(p_{\perp }-p_{\Vert })\right] \frac{(\mathbf{B}\cdot \nabla )%
\mathbf{B}}{4\pi }  \nonumber \\
&&\qquad +\mathbf{B}(\mathbf{B}\cdot \nabla )\left( \frac{p_{\perp
}-p_{\Vert }}{B^{2}}\right) =0.  \label{pressure_balance}
\end{eqnarray}%
Projection along the magnetic field gives 
\begin{equation}
-\nabla _{\parallel }p_{\Vert }-\frac{4\pi \left( p_{\perp }-p_{\Vert
}\right) }{B^{2}}\nabla _{\parallel }\frac{B^{2}}{8\pi }=0,
\label{parallel-projection}
\end{equation}%
which coincides with Eq. (9.2) of Shafranov' review \cite{shafranov-66}.  It
is possible to prove  that the solvability condition (\ref%
{parallel-projection}) reduces to an identity by means of \ both stationary
kinetic equations  (\ref{kineticEqs}) and the quasi-neutrality condition
(see, for instance,  \cite{NorthropWhiteman1964,  Grad1967,
HallMcNamara1975, ZakharovShafranov}).  Since the pressures depend on $B$
only, Eq. (\ref{parallel-projection}) reduces to  
\begin{equation}
-\frac{dp_{\Vert }}{dB}=\frac{\left( p_{\perp }-p_{\Vert }\right) }{B}.
\label{identity3}
\end{equation}

The existence of this identity means that for stationary states only two
scalar equations survive. Together with the condition $\nabla\cdot \mathbf{B}%
=0$, they provide a closed system of three equations for the three
components of the magnetic field.

Defining $\nabla _{\perp }=\nabla -B^{-2}\mathbf{B}(\mathbf{B}\cdot \nabla )$%
, the perpendicular component of Eq. (\ref{pressure_balance}) reads 
\begin{eqnarray}
&&-\nabla _{\perp }p_{\perp }+\frac{4\pi }{B^{2}}(p_{\perp }-p_{\Vert
})\nabla _{\perp }\frac{B^{2}}{8\pi }  \nonumber \\
&&\qquad +\left[ 1+\frac{4\pi }{B^{2}}(p_{\perp }-p_{\Vert })\right] \frac{%
\Big[ \left[ \nabla \times \mathbf{B}\right] \times \mathbf{B}\Big] }{4\pi }%
=0,  \label{perp-equilibrium}
\end{eqnarray}
which coincides with Eq. (9.3) of Shafranov's review \cite{shafranov-66}.

{\bf The two-dimensional problem}.
In two dimensions, we define the stream function $\psi $ (or vector
potential), such that $B_{x}={\partial \psi }/{\partial y}$, $B_{y}=-{%
\partial \psi }/{\partial x}$. In terms of $\psi $ and $B_{z}$, 
\begin{eqnarray}
&&\Big[ \Big[ \nabla \times \mathbf{B}\Big] \times \mathbf{B}\Big] =\mathbf{e%
}_{x}\Big( -\frac{1}{2}\frac{\partial B_{z}^{2}}{\partial x}-\frac{\partial
\psi }{\partial x}\Delta \psi \Big)  \nonumber \\
&&\qquad +\mathbf{e}_{y}\Big( -\frac{1}{2}\frac{\partial B_{z}^{2}}{\partial
y}-\frac{\partial \psi }{\partial y}\Delta \psi \Big) -\mathbf{e}_{z}\left\{
\psi ,B_{z}\right\},
\end{eqnarray}
where $\left\{ \psi ,B_{z}\right\}$ denotes the Jacobian. Furthermore, $%
\nabla _{\perp }=\nabla -\frac{1}{B^{2}}\mathbf{B}_{\perp }\mathbf{(B_{\perp
}\cdot \nabla )-}\frac{B_{z}}{B^{2}}\mathbf{e}_{z}(\mathbf{B}_{\perp }\cdot
\nabla )$, where $\nabla \equiv (\partial_x, \partial_y)$ and $\mathbf{B}%
_{\perp }=(B_{x},B_{y})$.

In Eq. (\ref{perp-equilibrium}), we now separate the $(x,y)$-components: 
\begin{eqnarray}  \label{perp}
&&-\nabla p_{\perp }+\frac{1}{B^{2}}\mathbf{B}_{\perp }\mathbf{(\mathbf{B}%
_{\perp }\cdot \nabla )}p_{\perp }  \nonumber \\
&& \quad+\frac{1}{2B^{2}}(p_{\perp }-p_{\Vert })\left[ \nabla -\frac{1}{B^{2}%
}\mathbf{B}_{\perp }\mathbf{(\mathbf{B}_{\perp }\cdot \nabla )}\right] B^{2}
\\
&&\quad +\frac{1}{4\pi }\left[ 1+\frac{4\pi }{B^{2}}(p_{\perp }-p_{\Vert })%
\right] \left( -\frac{1}{2}\nabla B_{z}^{2}-\nabla \psi \Delta \psi \right)
=0. \nonumber
\end{eqnarray}%
Equation for $z$ component, due to identity (\ref{identity3}), 
can be written as
\begin{eqnarray}
&&\frac{B_{z}}{4\pi }\left[ (\mathbf{B}_{\perp }\cdot \nabla )\left( 1+\frac{%
4\pi }{B^{2}}(p_{\perp }-p_{\Vert })\right) \right]  \nonumber \\
&&+\frac{1}{4\pi }\left[ 1+\frac{4\pi }{B^{2}}(p_{\perp }-p_{\Vert })\right]
(\mathbf{B}_{\perp }\cdot \nabla )B_{z}=0.
\end{eqnarray}
In terms of $\psi$, after integration, it leads to 
\begin{equation}
\frac{B_{z}}{4\pi }\left( 1+\frac{4\pi }{B^{2}}(p_{\perp }-p_{\Vert
})\right) =f(\psi ).  \label{B-z1}
\end{equation}%
Interestingly, in the isotropic case ($p_{\perp }-p_{\Vert }=0)$, we have $%
B_{z}=B_{z}(\psi ),$ in a full agreement with the Grad-Shafranov reduction 
\cite{grad,shafranov-58,shafranov-66}. Furthermore, because the projection
of the full equation on $\mathbf{B}$ is equal zero, in the 2D case where the
fields are functions of $x$ and $y$ only, the projection of Eq. (\ref{perp})
on $\mathbf{B_{\perp }}$ vanishes identically. Therefore the relevant
information is obtained by taking the vector product of Eq. (\ref{perp})
with $\mathbf{B_{\perp }}$, in the form 
\begin{eqnarray}
&& \Big(\nabla \psi \cdot \nabla ( p_{\perp}+\frac{B_{z}^{2}}{8\pi })\Big) 
  -\frac{(p_{\perp }-p_{\Vert })}{2B^{2}}\left( \nabla \psi \cdot \nabla
\left( B^{2}-B_{z}^{2}\right) \right)   \nonumber \\
&& \qquad =-\frac{\left( B^{2}-B_{z}^{2}\right) }{4\pi }\left[ 1+\frac{4\pi 
}{B^{2}}(p_{\perp }-p_{\Vert })\right] \Delta \psi.  \label{analog}
\end{eqnarray}%
This equation is supplemented by relation (\ref{B-z1}).

Equation (\ref{analog}) can be viewed as analogous to Grad-Shafranov
equation, the main difference being that the pressures are here prescribed
as functions of the magnetic field amplitude. Therefore, Eq. (\ref{analog})
does not reduce in the isotropic case to Grad-Shafranov equation, as seen
from identity (\ref{identity3}).

\section{\protect\bigskip Variational principle}

We now consider the purely two-dimensional geometry where $B_{z}=0$. In this
regime, $B^{2}=|\mathbf{B}_{\perp }|^{2}$ and Eq. (\ref{analog}) reduces to 
\begin{equation}
0=\frac{4\pi }{B^{2}}\Big[ \nabla \psi \cdot \Big( B\nabla \frac{p_{\perp }}{%
B}+\frac{p_{\Vert }}{B}\nabla B\Big) \Big] +\Big[ 1+\frac{4\pi }{B^{2}}%
(p_{\perp }-p_{\Vert })\Big] \Delta \psi.  \label{psi}
\end{equation}%
By means of Eq. (\ref{identity3}), it is easily checked that 
\[
\nabla \frac{4\pi }{B^{2}}(p_{\perp }-p_{\Vert })=\frac{4\pi }{B^{2}}\left(
B\nabla \frac{p_{\perp }}{B}+\frac{p_{\Vert }}{B}\nabla B\right). 
\]%
Therefore, Eq. (\ref{psi}) takes the form%
\begin{equation}
\nabla \cdot \left\{ \left[ 1+\frac{4\pi }{B^{2}}(p_{\perp }-p_{\Vert })%
\right] \nabla \psi \right\} =0  \label{Eq-psi}
\end{equation}%
and thus derives from the variational principle $\delta \mathcal{F}=0$ with $%
\mathcal{F}=\frac{1}{4\pi }\int g(|\nabla \psi |^{2})dxdy$. It rewrites 
$ \nabla g^{\prime }(B^{2})\nabla \psi =0$, 
where the function $g$ is found by integrating 
\[
g^{\prime }(B^{2})=1+\frac{4\pi }{B^{2}}(p_{\perp }-p_{\Vert }). 
\]%
Due to identity (\ref{identity3}), we have 
\begin{equation}
\mathcal{F}=\int \left( \frac{B^{2}}{8\pi }-p_{\Vert }\right) \, dx \, dy
\equiv -\int \Pi_\| \, dx \, dy .  \label{free-general}
\end{equation}%
It follows that all the two-dimensional stationary states in anisotropic MHD
are stationary points of the functional $\mathcal{F}$. Its density is a
function of $B=|\nabla \psi |$ only. In the special case of cold electrons,
this free energy turns out to identify with the Hamiltonian of the static
problem \cite{PRS06}.

Equations similar to (\ref{Eq-psi}) arise in the context of pattern
structures in thermal convection. As shown in \cite%
{ErcolaniIndikNewellPassot}, such equations represent integrable
hydrodynamic systems. As in the usual one-dimensional gas dynamics, these
systems display breaking phenomena where the solution looses its smoothness
at finite distance, due to the formation of folds. As a consequence, these
models require some regularization. For patterns, the authors of \cite%
{ErcolaniIndikNewellPassot} supplement in the equation an additional linear
term involving a square laplacian. In our case, this procedure corresponds
to the replacement of $\mathcal{F}$ by $\mathcal{F}+ ({\nu}/{2})\int \left(
\Delta \psi \right)^{2}dxdy$, with a constant $\nu >0$. In plasma physics,
regularization can originate from finite Larmor radius (FLR) corrections,
which are not retained in the present analysis based on the drift kinetic
equation (see, e.g. \cite{KPS07a, KPS07b}).

\bigskip

\section{Adiabatic approximation}

{\bf Equations of state and their regularization}.
To specify the model, we consider stationary mirror structures which result
from the non-linear development of the mirror instability (MI), one of the
slowest instabilities in plasma physics. The characteristic frequencies of
mirror modes are much smaller the ion gyro-frequency, which suggests to use
(at least at sufficiently large scales) a description based on the drift
approximation for the particle distribution functions which, for stationary
states, reduces to Eq. (\ref{kineticEqs}).

In order to specify these distribution functions, we need to connect the
initial state (where both ions and electrons species are assumed
biMaxwellian) with the stationary distribution functions. In the weakly
nonlinear regime that develops near threshold, the transition from the
initial homogeneous state to the weakly nonlinear one is slow in time, so
that, to leading order, the distribution function $f_{\alpha }$ as a
function of $\mu $ and $W_{\alpha }$ retains its form during the evolution 
\cite{KPS12}. Therefore, the function $f_{\alpha }(\mu ,W_{\alpha })$ can be
determined by matching with the initial distribution function 
$
f_{\alpha }^{(0)}=A_{\alpha }\exp \big[-\frac{v_{\parallel }^{2}}{%
v_{\parallel \alpha }^{2}}-\frac{\mu B_{0}m_{\alpha }}{T_{\perp \alpha }}%
\big],  
$
which corresponds to $\phi =0$ and $W_{\alpha }=\frac{v_{\Vert }^{2}}{2}+\mu
B_{0}$, where $A_{\alpha }~=~n_{0}m_{\alpha }/(2\pi \sqrt{\pi }v_{\parallel
\alpha }T_{\perp \alpha })$. Here $T_{\parallel \alpha }$ and $T_{\perp
\alpha }$ are the initial perpendicular and transverse temperatures, $B_{0}$
the initial homogeneous magnetic field, and $v_{\parallel \alpha }=\left(
2T_{\parallel \alpha }/m_{\alpha }\right) ^{1/2}$ the parallel thermal
velocity. As a result of the matching, we get \cite{KPS12} 
\begin{equation}
f_{\alpha }(\mu ,W_{\alpha })
=A_{\alpha }\exp \Big [-\frac{2W_{\alpha }}{v_{\parallel \alpha }^{2}%
}+\mu B_{0}m_{\alpha }\Big(\frac{1}{T_{\parallel \alpha }}-\frac{1}{T_{\perp
\alpha }}\Big)\Big].  \label{g-alpha}
\end{equation}%
Note that this function has the Boltzmann form with respect to $W_{\alpha }$
but display, at fixed $W_{\alpha }$, an exponential growth relatively to $%
\mu $ when $T_{\perp \alpha }> T_{\parallel \alpha }$, a necessary condition
for MI. Such a growth, however, leads to a singular behavior of the
pressures as functions of $B$. Indeed, for the distribution function (\ref%
{g-alpha}), the parallel pressure is \cite{KPS12}:%
\begin{equation}
p_{\Vert }=n_{0}(T_{\Vert i}+T_{\Vert e})\frac{1+u}{\left( 1+a_{e}u\right)
^{c_{e}}\left( 1+a_{i}u\right) ^{c_{i}}},
\end{equation}%
where $u=B/B_{0}-1$, $a_{\alpha }=T_{\perp \alpha }/T_{\parallel \alpha }$
is the parameter characterizing the anisotropy of distribution function $%
f_{\alpha }$, and $c_{\alpha }=T_{\parallel \alpha }(T_{\parallel
e}+T_{\parallel i})^{-1}$ in the case of a proton-electron plasma. The
singularities at $u=-a_{\alpha }^{-1}$ correspond to the magnetic field 
\begin{equation}
B_{s}=B_{0}\frac{a_{\alpha }-1}{a_{\alpha }}<B_{0}.  \label{singularB}
\end{equation}
In the limiting case of cold electrons, $p_{\Vert }=n_{0}T_{\Vert
}(1+u)(1+au)^{-1}$ displays a pole singularity. Here, $T_{\Vert }$ and the
anisotropy parameter $a$ correspond to ions only. Such an equation of state
was previously derived by a quasi-normal closure of the fluid hierarchy \cite%
{PRS06}.

The above singularities are presumably related to an overestimated
contribution from large $\mu $, corresponding either to small $B$ or to
large a transverse kinetic energy. In both cases, the applicability of the
drift approximation breaks down and we are thus led to introduce some
cut-off type correction near $\mu_\alpha ^{\ast }$. In a simple variant, we
take $f_{\alpha }=\tilde C_\alpha\exp(-m_\alpha W_\alpha/T_{\parallel\alpha})
$ at $\mu>\mu_\alpha ^{\ast }$, with some positive constant $\tilde C_\alpha$%
, and $f_{\alpha}$ retains its original form (\ref{g-alpha}) for $\mu \leq
\mu_\alpha ^{\ast }$. For cold electrons case, the parallel ion pressure is
modified into $p_{\Vert }=n_{0}T_{\Vert }G(B,r)$ with 
\[
G(B,r)=\frac{1}{1+C}\left[\frac{(B_{0}-B_{s})B}{B_{0}(B-B_{s})}R(B,r)
+Ce^{r(B_0-B)}\right], 
\]
where 
\[
R(B,r)=\frac{\exp [-r(B-B_{s})]-1}{\exp [-r(B_{0}-B_{s})]-1}, 
\]
$C$ is a (small) constant, and $r=m\mu ^{\ast }/T_{\Vert }$. Noticeably,
regularization leads to a non-singular positive pressure for all $B$,
including when $B\rightarrow 0$. The modification for $p_{\Vert }$ in the
case of hot electrons is not specified here because the expressions are
algebraically much more cumbersome but do not involve any additional
difficulty.

{\bf KP soliton}.
We now show that the functional $\mathcal{F}$ we previously introduced has
the meaning of a free energy. In the weakly nonlinear regime near the MI
threshold, the temporal behavior of the mirror modes can be described by a
3D model \cite{KPS07a, KPS07b, KPS12}, that in the present 2D geometry reads 
\begin{equation}
u_{t}=-\widehat{|k_{y}|}\frac{\delta F}{\delta u}  \label{asymp}
\end{equation}
with the free energy 
\begin{equation}
F=\int \left[ \frac{1}{2}(-\varepsilon u^{2}+u\frac{\partial _{z}^{2}}{%
\Delta _{\perp }}u+\left( \nabla _{\perp }u\right) ^{2})+\frac{\lambda }{3}%
u^{3}\right] d\mathbf{r.}  \label{F-3D}
\end{equation}%
Here $u$ denotes the dimensionless magnetic field fluctuations and $%
\varepsilon$ the distance from MI threshold. The third term in $F$
originates for the FLR corrections, and $\lambda$ is a nonlinear coupling
coefficient which is positive for bi-Maxwellian distributions. In Eq. (\ref%
{asymp}), the operator $\widehat{|k_{y}|}$ is a positive definite operator
(in the Fourier representation it reduces to $|k_{y}|$), so that Eq. (\ref%
{asymp}) has a generalized gradient form.

Let us now show that this result can be obtained from the functional $%
\mathcal{F}$ defined in (\ref{free-general}). We isolate the perturbation $%
\varphi $ in the stream function $\psi =-B_{0}(x+\varphi )$ with $\varphi
\rightarrow 0$ as $|\mathbf{r}|\rightarrow \infty $, so that the mean
magnetic field $\mathbf{B}_{0}$ is directed along the $y$-axis. We then
expand Eq. (\ref{free-general}) in series with respect to $u$. For the sake
of simplicity, we restrict the analysis to the case of cold electrons. The
expansion of the integrand $B^{2}/(8\pi) -p_{\Vert }$ in $F$ has then the
form 
\begin{eqnarray}
&&n_{0}T_{\parallel }\left[ \frac{\left( u+1\right) ^{2}}{\beta _{\Vert }}-%
\frac{1+u}{1+au}\right]  \nonumber \\
&&\qquad =n_{0}T_{\parallel }[ \left( \beta _{\Vert }^{-1}-1\right) +u\left(
a+2\beta _{\Vert }^{-1}-1\right)  \nonumber \\
&&\qquad +u^{2}\left( -a^{2}+a+\beta _{\Vert }^{-1}\right) -u^{3}
a^{2}\left( a-1\right) +....]  \label{expansion}
\end{eqnarray}
where we use the usual notation $\beta _{\Vert }=8\pi
n_{0}T_{\parallel}/B_{0}^{2}$.

As well known (see, e.g. \cite{KPS07a, KPS07b}), near threshold, MI develops
in quasi-transverse directions relative to $\mathbf{B}_{0}$. This means
that, in the 2D geometry, $\varphi _{x}\gg \varphi _{y}$ and, with a good
accuracy, $u$ coincides with $\varphi _{x}$. However, in the expansion of $u=%
\sqrt{(\varphi _{x}+1)^{2}+\varphi _{y}^{2}}-1 \simeq \varphi _{x}+
\varphi_{y}^{2}/{2}$, it is necessary to keep the second term, quadratic
with respect to $\varphi $. The linear term in expansion of $\mathcal{F}$
vanishes and the quadratic terms is given by 
\[
\mathcal{F}_{2}=n_{0}T_{\parallel }\int \Big\{ \Big[ a(a-1)+\frac{1}{\beta}
_{\Vert }\Big] \varphi _{x}^{2}+\Big[ a-1+\frac{2}{\beta} _{\Vert }\Big ]
\varphi _{y}^{2}\Big \} dxdy. 
\]%
where the factor $a(a-1)+1/\beta \equiv -\varepsilon /2$ defines the MI
threshold $a=1+{1}/{\beta _{\perp }}$ (that the present equations of state
accurately reprocuces). It is also seen that for $|\varepsilon| \ll 1$, $%
\varphi _{x}/\varphi _{y}\sim |\varepsilon |^{-1/2}$, in agreement with the
quasi-one-dimensional development of MI near threshold. In this case, $%
\mathcal{F}_{2}$ coincides with the quadratic term in (\ref{F-3D}), up to a
simple rescaling and to the FLR contribution, Furthermore, the cubic term in
(\ref{expansion}) gives the nonlinear coupling coefficient $\lambda =
a\left( a-1\right) >0$. As a consequence, $\mathcal{F}$, introduced in the
previous section, reduces to the free energy of the asymptotic model. The
temporal equation for $\varphi $ has also the generalized gradient form
originating from (\ref{asymp}), 
\begin{equation}
\varphi _{t}=-\Gamma \frac{\delta F}{\delta \varphi} \ \text{with} \ \Gamma
= -\widehat {\frac{|k_{y}|}{k_{x}^{2}}},  \label{phi-F}
\end{equation}
for which the associated stationary equation reads 
\begin{equation}
\varepsilon \varphi _{xx}+\varphi _{xxxx}-\varphi _{yy}-\lambda \partial
_{x}\left( \varphi _{x}^{2}\right) =0,  \label{stat-phi}
\end{equation}%
where the linear operator $L=-\varepsilon \partial _{xx}+\partial
_{yy}-\partial _{xxxx}$ is elliptic or hyperbolic depending on the sign of $%
\varepsilon $. For $\varepsilon >0$ (above threshold), this operator is
hyperbolic, while below threshold it is elliptic and thus invertible in the
class of functions vanishing at infinity. Remarkably, in the latter case,
Eq. (\ref{stat-phi}) identifies with the soliton for KP equation called
lump. In standard notations, lump is indeed a solution of the stationary
KP-II equation, 
\begin{equation}
-Vu_{xx}+u_{xxxx}-u_{yy}+3(u^{2})_{xx}=0,  \label{KP}
\end{equation}%
where $V$ is the lump velocity. When comparing this equation with (\ref%
{stat-phi}) we see that $-|\varepsilon |$ plays the role of the lump
velocity $V $ and $\lambda \varphi _{x}\rightarrow -3u$.

The lump solution was first discovered numerically by Petviashvili \cite%
{Petviashvili} using the method now known as the Petviashvili scheme (see
the next section). The analytical solution was later on obtained in \cite%
{BIMMZ}. In our notation, it reads%
\[
\varphi _{x}=-\frac{12|\varepsilon |}{\lambda }\frac{\left( 3+ \varepsilon
^{2}y^{2}-|\varepsilon |x^{2}\right) }{\left[ 3+ \varepsilon
^{2}y^{2}+|\varepsilon |x^{2}\right] ^{2}}. 
\]%
This function vanishes algebraically at the infinity like $r^{-2}$. In the
center region $-|\varepsilon |^{-2}\sqrt{|\varepsilon |x^{2}-3}%
<y<|\varepsilon |^{-2}\sqrt{|\varepsilon |x^{2}-3}$, the magnetic field
displays a hole with a minimum at $x=y=0$ equal to $-4|\varepsilon |/\lambda$%
. In the outer region, the magnetic lump has two symmetric humps with
maximum values $|\varepsilon|/(2\lambda)$ at $y=0$ and $x=\pm 3|\varepsilon
|^{-1/2}$. The main contribution to the ``skewness'' $I=\int \varphi _{x}
^{3} \,dx\,dy$ comes from the hole region, providing a negative value to $I$%
, in complete agreement with \cite{KPS07a, KPS07b}.

{\bf Numerical solutions -- the methods}.
In the 2D case, our regularized model equation 
for stationary pressure-balanced structures has a variational form 
\begin{equation}
-\partial _{x}\left[ \frac{(1+\varphi _{x})}{(1+u)}\frac{dg}{du}\right]
-\partial _{y} \left[ \frac{\varphi _{y}}{(1+u)}\frac{dg}{du}\right]+\nu
\Delta ^{2}\varphi =0.  \label{variational_form_reg}
\end{equation}
Clearly, Eq. (\ref{variational_form_reg}) describes stationary points 
$\delta \mathcal{F}/\delta \varphi=0$ of the functional 
$\mathcal{F}=\int [g(u)+(\nu/2) (\Delta \varphi)^{2}]\,dx\,dy$, 
with some constant parameter $\nu $. (In this expression and everywhere
below we use dimensionless variables). 

We applied two numerical methods to solve Eq. (\ref{variational_form_reg}).
The first one is a generalization of the well known gradient method which
corresponds to a dissipative dynamics along an auxiliary time-like variable $%
\tau $ of the form $\varphi _{\tau }=-\widehat{\Gamma} ({\delta {\mathcal{F}}%
}/{\delta \varphi})$, with a positively definite linear operator $\widehat{%
\Gamma}$. It is clear that attractors in the phase space of the above
dynamical system are stable solutions of Eq. (\ref{variational_form_reg}).
Unstable solutions however cannot be found by this method.

Furthermore, the linear part of Eq. (\ref{variational_form_reg}) is of the
form ${\widehat L} \varphi= -g^{\prime \prime }(0) \varphi _{xx}-g^{\prime
}(0)\varphi _{yy}+\nu \Delta^2 \varphi$. The coefficient $g^{\prime \prime
}(0)$ is proportional to $\varepsilon$ (introduced in the previous section)
and $g^{\prime }(0)$ is positive within the adiabatic approximation. When
these two are positive, the operator $L$ is elliptic and it is possible to
employ the so-called Petviashvili method \cite{Petviashvili}. It is a
specific method for finding localized solutions of equations of the form 
$
\widehat{M}\varphi =N[\varphi],
$
with a positively definite linear operator $\widehat{M}$ and a nonlinear
part $N[\varphi ]$. Note that in our case the Fourier image of $\widehat{M}$
is 
\begin{equation}
M(k_{x},k_{y})= g^{\prime \prime }(0)k_{x}^{2}+g^{\prime }(0)k_{y}^{2}+\nu
(k_{x}^{2}+k_{y}^{2})^{2}>0.
\end{equation}%
In its simplest form, the iteration scheme of the Petviashvili method reads 
\begin{equation}
\varphi _{n+1}=(\widehat{M}^{-1}N[\varphi _{n}])\left( \frac{\int \varphi
_{n}\widehat{M}\varphi _{n}\,dx\,dy}{\int \varphi _{n}N[\varphi _{n}]\,dx\,dy%
}\right) ^{-\gamma },
\end{equation}%
where $\gamma $ is a positive parameter in the range $1<\gamma <2$. The
corresponding multiplier strongly affects the structure of attractive
regions in the phase space.

It is worth noting that if the operator ${\widehat L}$ is hyperbolic,
solutions of the problem are not localized with respect to both $x$ and $y$
coordinates, and will be periodic or more generally quasiperiodic \cite{ZK,
KD}.

\begin{figure}[tbp]
\begin{center}
\includegraphics[width=0.45\textwidth]{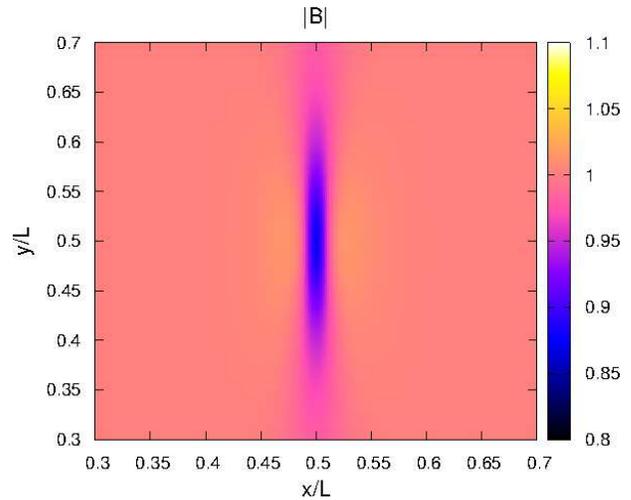}
\end{center}
\caption{Fig. 1. Unstable localized solution for $\protect\nu=0.0004$, $r=7$%
, $B_{s}=0.5$ (in units of $B_{0}$), and $C=0.002$. The value $1/\protect%
\beta_\parallel=1.127$ prescribes an aspect ratio $\protect\sqrt{g^{\prime
\prime }(0)/g^{\prime }(0)}=0.2$.}
\label{lump}
\end{figure}

\begin{figure}[tbp]
\begin{center}
\includegraphics[width=0.4\textwidth]{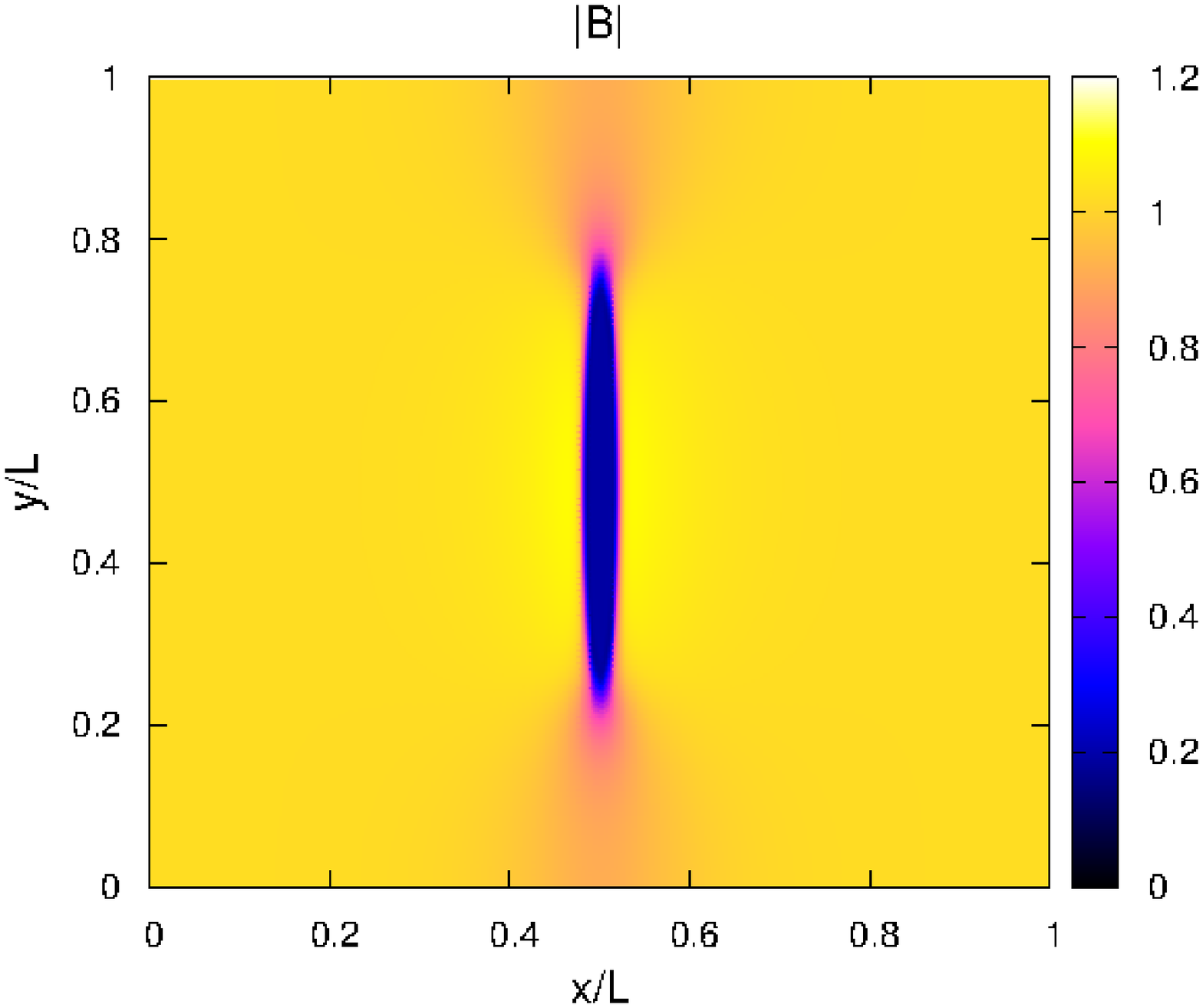}\\[0pt]
\vspace{4mm} \includegraphics[width=0.4\textwidth]{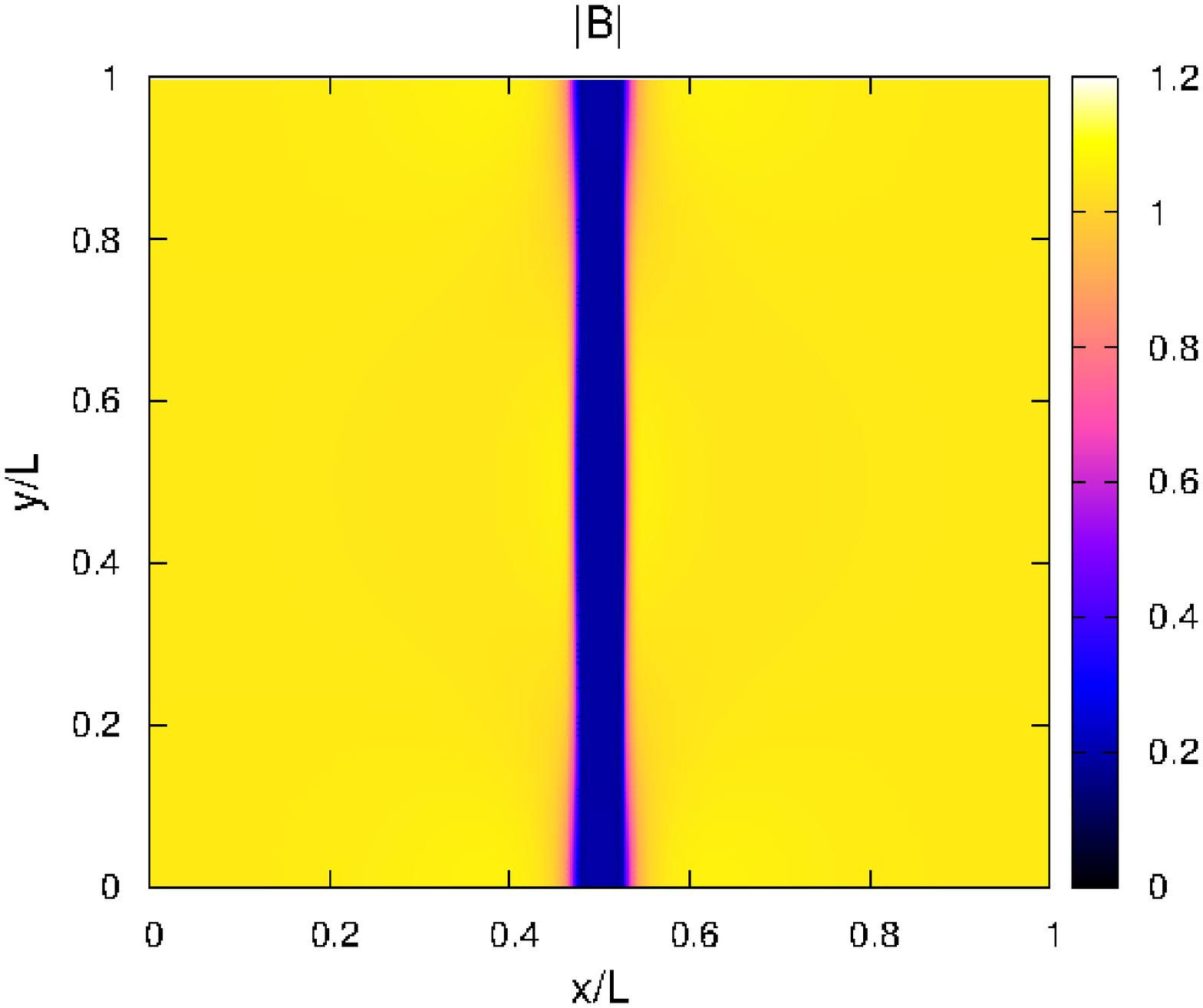}\\[0pt]
\vspace{4mm} \includegraphics[width=0.4\textwidth]{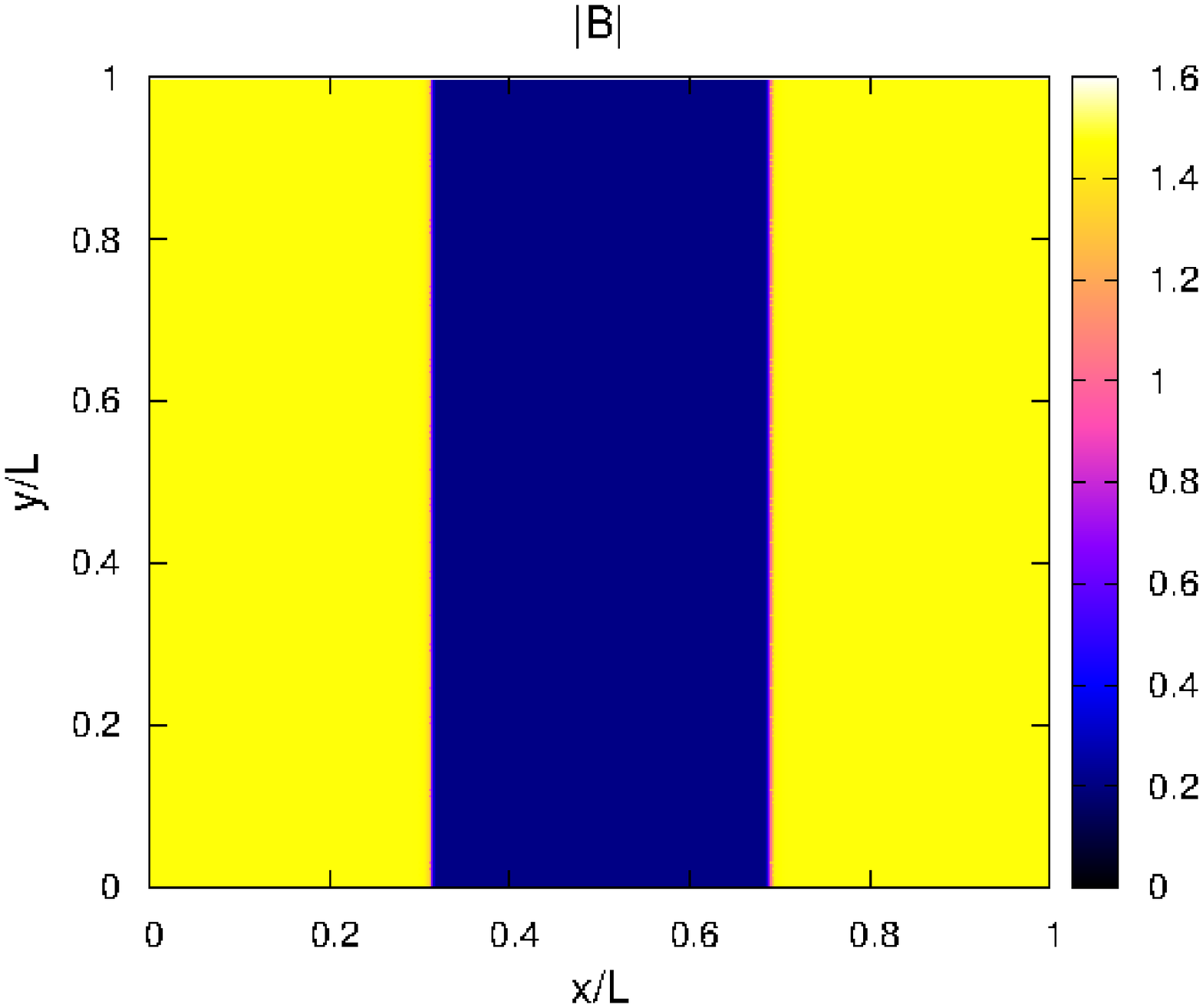}
\end{center}
\caption{Fig. 2. Formation of a stable 1D solution in a gradient
computation, for the same parameters as in Fig.1.}
\label{stripe}
\end{figure}

{\bf The results}. 
We performed computations with both
numerical methods using fast Fourier transform numerical routines for the
evaluation of the linear operators. Periodic boundary conditions for a
computational square $2\pi\times 2\pi$ were assumed.

For the gradient method, we used the simplest first-order Euler scheme for
stepping along $\tau$, with $\delta\tau\sim 0.01 $.  The operator $\widehat
\Gamma$ was taken in a form giving stable computation, namely $%
\Gamma(k_x,k_y)=1/[k_x^2+k_y^2+ \nu(k_x^2+k_y^2)^2]$.

As for the Petviashvili method, the value $\gamma=1.8$ was used, leading,
after an erratic transient, to a convergence of the iterations to unstable
solutions of the variational equation (\ref{variational_form_reg}).

The main results of our computations can be formulated as follows. There do
exist unstable localized solutions of Eq. (\ref{variational_form_reg}),
which are similar to the lump solutions of KPII equation, when written in
terms of $u=\partial_x\varphi$ (Fig. \ref{lump}). For asymptotically 
small $\varepsilon$, they accurately coincide with KP solutions, 
independently of the electron temperature, as it should be. 
Such low-amplitude stationary states do not depend on the particular 
choice of the regularization of $g(u)$. No other kinds of solutions were 
found with the Petviashvili method.

When the gradient method is used, large amplitudes $u\sim 1$ are achieved in
many cases, and the final result turns out to be dependent on the choice of
the parameters $r$ and $C$ in the regularized function $g$. Without
regularization, no smooth stationary state is approached. Instead, a
singularity occurs. Differently, when a regularized $g$ with parameters $%
r\sim 10$ and $C\sim 0.001$ is used, the final state identifies with a
one-dimensional stripe in the form of a magnetic hole, as shown 
in Fig. \ref{stripe} that also displays typical stages of the ``gradient'' 
evolution. In all simulations, the magnetic field in the stripe was smaller 
than the `singular' magnetic field $B_{s}$ given by Eq. (\ref{singularB}). For
increasing $r$, the magnetic field in the stripe tends to decrease, down to
0. For initial conditions in the form of a slightly perturbed 2D lump, the
final result is always a one-dimensional stripe of hole type, which
demonstrates the instability of the 2D lump, in full agreement with the
analytical prediction \cite{KPS07a, KPS07b}.

In no cases stable 2D structures localized both in $x$ and $y$ directions
were found. Instead, the gradient method showed that stable structures can
only be one-dimensional, transverse to the magnetic field. An initial
localized perturbation of sufficiently high amplitude develops into an
increasingly long structure along the $y$ axis, and eventually reaches the
boundary of the computational domain.

The question arises whether the 1D shock solutions obtained in \cite{PRS06}
(for which $\mathrm{min} B >B_s$) would identify with the present solution
when $\nu \to 0$, a limit which is unreachable in the present numerics. It
is possible that the presence of the bi-Laplacian regularization leads to
overshooting in the shock solution, resulting in the convergence towards
solutions where $\mathrm{min} B < B_s$.

\section{Conclusion}

A detailed analysis was presented for the Grad-Shafranov equations
describing static force-balanced mirror structures with anisotropic
pressures given by equations of state derived from drift kinetic equations,
when assuming an adiabatic evolution from bi-Maxwellian initial conditions.
It turns out that in two dimensions, the problem is amenable to a
variational formulation with a free energy provided by the space integral of
the parallel tension. Slightly below the mirror instability threshold, small
amplitude solutions associated to KPII lumps are obtained and shown to be
unstable. Differently, when considering stable subcritical structures, the
drift kinetic approximation breaks down, as the deep magnetic holes obtained
by a gradient method appear to be strongly sensitive to the regularization
process, an effect which in a more realistic description could be provided
by FLR corrections and/or particle trapping.

The authors thank the referees for valubable remarks. This work was
supported by CNRS PICS programme 6073 and RFBR grant 12-02-91062-CNRS-a.
T.P. and P.L.S. benefited from support from INSU-CNRS PNST. The work of E.K.
and V.R. was also supported by the RAS Presidium Program "Fundamental
problems of nonlinear dynamics in mathematical and physical sciences" and
Grant NSh 6170.2012.2.

\end{document}